\documentclass[aps,preprint]{revtex4}%
\usepackage{amssymb}
\usepackage{amsfonts}
\usepackage{amsmath}
\usepackage{graphicx}%
\providecommand{\U}[1]{\protect\rule{.1in}{.1in}}

\begin{document}
\title{Fermi coordinates and static observer in Schwarzschild spacetime. \ }
\author{V. A. Belinski}
\affiliation{Institut des Hautes Etudes Scientifiques (IHES), F-91440, Bures-sur-Yvette, France,}
\affiliation{Universit\'{e} Libre de Bruxelles and International Solvay Institutes, B-1050
Brussels, Belgium}
\affiliation{International Network of Centers for Relativistic Astrophysics (ICRANet),
65122 Pescara, Italy.}

\begin{abstract}
In this paper we construct the Fermi coordinates along any arbitrary line in
simple analytical way without use orthogonal frames and their transport. In
this manner we extend the Eddington approach to the construction of the Fermi
metric in terms of the Riemann tensor.\ In the second part of the present
article we show how the proposed approach works practically by applying it for
deriving the Fermi coordinates for the static observer in the Schwarzschild spacetime.

\end{abstract}
\maketitle

\section{Introduction}

It is known that for any metric and any line exists a set of Fermi coordinates
\cite{Fe} in which all Christoffel symbols are zero at points of this line and
\textit{this is the definition of Fermi coordinates}. However, the elimination
of the Christoffel symbols on a line does not fix completely the corresponding
coordinate transformations which means that there is an infinity of the Fermi
coordinates associated to a given line. To make a concrete choice it is
reasonable to search for some additional coordinate restrictions (not
violating the vanishing of Christoffel symbols on line) appropriate from a
physical point of view. The natural physical support have been proposed by
Arthur Eddington \cite{Ed} who also developed the way for the corresponding
analytical calculations. Eddington did this for the case of the Riemann
coordinates in the neighborhood of a point in 4-dimensional spacetime\textit{
}(by definition in Riemann coordinates all Christoffel symbols are zero at
some point of spacetime and not along a line). His idea was to specify the
coordinate transformations so as to represent the quadratic terms of the
expansion of the metric near such point by the components of the Riemann
tensor. It turn out that the generalization of Eddington approach to the case
of Fermi coordinates in the neighborhood of an arbitrary line is
straightforward. Such extension is the target of the first part of the present
paper. It should be stressed that it is done for any\textit{\ }original metric
and any given curve, no matter what is its geometric character (geodesic or
not, timelike, spacelike or null) and in pure analytical way without necessity
to use orthogonal frames and their Fermi-Walker transport. Such simplified
universal method has some value, because the majority of papers in the
literature have been dedicated only to some specific type of the line and have
been essentially based on the use of transported frames (for example, Manasse
and Misner \cite{MM} did this for timelike geodesics and Blau, Frank and Weiss
\cite{Bl} extended their results for null geodesic curve).

In the second part of the present article we show the proposed approach in
practical action by applying it for construction of the Fermi coordinates for
the static observer in the Schwarzschild spacetime. This result is new since
the known analogous constructions (for example, see \cite{BGJ}\ and references
therein) have been restricted to a quasi-Fermi system defined by Synge
\cite{Sy} when not all Christoffel symbols on the world line of interest disappear.

It is worth to remark that since the work of Synge some terminological muddle
has been widely spread in the literature. Synge introduced coordinates which
he named "Fermi coordinates" in spite of the fact that in general this
contradicts to the generally accepted understanding of what are Fermi
coordinates (which disparity was noted by Synge himself in his publication).
The Synge and Fermi coordinates coincide only for geodesic world lines but for
non-geodesics no Fermi coordinates can be constructed by the Synge
prescription. In general this prescription lead to the non-zero values of
Christoffel symbols $\Gamma_{0\alpha}^{0}$ and $\Gamma_{00}^{\alpha}$ at
points of a line and this is the reason to attribute to the Synge approach the
aforementioned "quasi-Fermi" appellation \cite{SS1}.

The problem is that for a non-geodesic line the Synge coordinate system is
essentially different from the Fermi coordinates and no article is known where
the Fermi coordinates would be constructed along non-geodesic line. The
present paper set aside to remove this shortage of traditional activity in
this field.

\section{Construction of Fermi coordinates in general}

It is known that for any metric $g_{ik}(x)$ (by symbol $x$ we denote the set
of 4 coordinates $x^{0},x^{1},x^{2},x^{3}$) in 4-dimensional spacetime
\cite{SS2} and any line%
\begin{equation}
x^{\alpha}=f^{\alpha}(x^{0}) \label{A1}%
\end{equation}
exists a set of Fermi coordinates $\acute{x}$ (that is $\acute{x}^{0}%
,\acute{x}^{1},\acute{x}^{2},\acute{x}^{3}$) in which all Christoffel
$\acute{\Gamma}$-symbols are zero at points of this line. For the
corresponding coordinates transformation $\acute{x}^{i}=\acute{x}^{i}%
(x^{0},x^{1},x^{2},x^{3})$ we denote the Jacobi matrix by $A_{k}^{i}$:%
\begin{equation}
A_{k}^{i}{}(x)=\frac{\partial\acute{x}^{i}}{\partial x^{k}}\text{ }.
\label{A2}%
\end{equation}
The transformation of $\Gamma$-symbols can be written as%
\begin{equation}
\Gamma_{kl}^{i}A_{i}^{q}=\acute{\Gamma}_{nm}^{q}A_{k}^{n}A_{l}^{m}+A_{k,l}%
^{q}\text{ }. \label{A3}%
\end{equation}
From the last formula follows that $\acute{\Gamma}_{nm}^{q}$ in Fermi
coordinates vanish on the line (\ref{A1}) if matrix $A_{k}^{i}$ satisfy the
differential equation:
\begin{equation}
\left[  A_{k,l}^{i}\right]  _{\mathcal{L}}=\left[  \Gamma_{kl}^{m}A_{m}%
^{i}\right]  _{\mathcal{L}}\text{ },\text{ } \label{A4}%
\end{equation}
where $\left[  F\right]  _{\mathcal{L}}$ means the value of any function $F$
on the line (\ref{A1}), that is%
\begin{equation}
\left[  F(x^{0},x^{1},x^{2},x^{3})\right]  _{\mathcal{L}}=F\left[  x^{0}%
,f^{1}(x^{0}),f^{2}(x^{0}),f^{3}(x^{0})\right]  )\text{ }. \label{A4-1}%
\end{equation}
It is easy to see that equation (\ref{A4}) represents the set of
ordinary\textit{\ }differential equations with respect to the variable $x^{0}%
$. Indeed, in the vicinity of the line (\ref{A1}) the transformation between
Fermi and original coordinates can be represented in form of an expansion with
respect to the three small deviations $x^{\alpha}-f^{\alpha}(x^{0})$ from the
line:%
\begin{gather}
\acute{x}^{m}=X^{m}(x^{0})+Y_{\alpha}^{m}(x^{0})\left[  x^{\alpha}-f^{\alpha
}(x^{0})\right] \label{A5}\\
+Z_{\alpha\beta}^{m}(x^{0})\left[  x^{\alpha}-f^{\alpha}(x^{0})\right]
\left[  x^{\beta}-f^{\beta}(x^{0})\right]  +O(3)\text{ },\nonumber
\end{gather}

where $O(n)$ means collection of terms of the order $n$ and higher with
respect to the small functional parameters $x^{\alpha}-f^{\alpha}(x^{0})$.
From (\ref{A5}) and definition (\ref{A2}) follow expansion for the components
of matrix $A_{k}^{m}$:%
\begin{equation}
A_{0}^{m}=\frac{dX^{m}}{dx^{0}}-Y_{\alpha}^{m}\frac{df^{\alpha}}{dx^{0}%
}+\left(  \frac{dY_{\beta}^{m}}{dx^{0}}-2Z_{\alpha\beta}^{m}\frac{df^{\alpha}%
}{dx^{0}}\right)  \left(  x^{\beta}-f^{\beta}\right)  +O(2)\text{ },
\label{A6}%
\end{equation}%
\begin{equation}
A_{\alpha}^{m}=Y_{\alpha}^{m}+2Z_{\alpha\beta}^{m}\left(  x^{\beta}-f^{\beta
}\right)  +O(2)\text{ }. \label{A7}%
\end{equation}
Consequently on the line the components $A_{k}^{m}$ are:%
\begin{equation}
\left[  A_{0}^{m}\right]  _{\mathcal{L}}=\frac{dX^{m}}{dx^{0}}-Y_{\alpha}%
^{m}\frac{df^{\alpha}}{dx^{0}}, \label{A8}%
\end{equation}%
\begin{equation}
\left[  A_{\beta}^{m}\right]  _{\mathcal{L}}=Y_{\beta}^{m}\text{ }. \label{A9}%
\end{equation}
From (\ref{A6}) and (\ref{A7}) follow values of the partial derivatives
$A_{k,l}^{m}$ of matrix $A_{k}^{m}$ on line:
\begin{equation}
\left[  A_{0,0}^{m}\right]  _{\mathcal{L}}=\frac{d}{dx^{0}}\left(
\frac{dX^{m}}{dx^{0}}-Y_{\alpha}^{m}\frac{df^{\alpha}}{dx^{0}}\right)
-\left(  \frac{dY_{\beta}^{m}}{dx^{0}}-2Z_{\alpha\beta}^{m}\frac{df^{\alpha}%
}{dx^{0}}\right)  \frac{df^{\beta}}{dx^{0}}\text{ }, \label{A10}%
\end{equation}%
\begin{equation}
\left[  A_{0,\beta}^{m}\right]  _{\mathcal{L}}=\frac{dY_{\beta}^{m}}{dx^{0}%
}-2Z_{\alpha\beta}^{m}\frac{df^{\alpha}}{dx^{0}}\text{ }, \label{A11}%
\end{equation}%
\begin{equation}
\left[  A_{\beta,0}^{m}\right]  _{\mathcal{L}}=\frac{dY_{\beta}^{m}}{dx^{0}%
}-2Z_{\alpha\beta}^{m}\frac{df^{\alpha}}{dx^{0}}\text{ }, \label{A12}%
\end{equation}%
\begin{equation}
\left[  A_{\alpha,\beta}^{m}\right]  _{\mathcal{L}}=2Z_{\alpha\beta}^{m}\text{
}. \label{A13}%
\end{equation}
It is convenient to use for the quantity $\left[  A_{0}^{m}\right]
_{\mathcal{L}}$ from (\ref{A8}) the special notation $\Lambda^{m}$:%
\begin{equation}
\Lambda^{m}=\frac{dX^{m}}{dx^{0}}-Y_{\alpha}^{m}\frac{df^{\alpha}}{dx^{0}%
}\text{ }. \label{A14}%
\end{equation}
After substitution expressions (\ref{A8})-(\ref{A13}) into equation (\ref{A4})
we find that this equation is equivalent to the following system:
\begin{equation}
\frac{d\Lambda^{m}}{dx^{0}}=\left\{  \left[  \Gamma_{\beta0}^{0}\right]
_{\mathcal{L}}\frac{df^{\beta}}{dx^{0}}+\left[  \Gamma_{00}^{0}\right]
_{\mathcal{L}}\right\}  \Lambda^{m}+\left\{  \left[  \Gamma_{\beta0}^{\alpha
}\right]  _{\mathcal{L}}\frac{df^{\beta}}{dx^{0}}+\left[  \Gamma_{00}^{\alpha
}\right]  _{\mathcal{L}}\right\}  Y_{\alpha}^{m}\text{ }, \label{A15}%
\end{equation}%
\begin{equation}
\frac{dY_{\beta}^{m}}{dx^{0}}=\left\{  \left[  \Gamma_{\alpha\beta}%
^{0}\right]  _{\mathcal{L}}\frac{df^{\alpha}}{dx^{0}}+\left[  \Gamma_{\beta
0}^{0}\right]  _{\mathcal{L}}\right\}  \Lambda^{m}+\left\{  \left[
\Gamma_{\alpha\beta}^{\gamma}\right]  _{\mathcal{L}}\frac{df^{\alpha}}{dx^{0}%
}+\left[  \Gamma_{\beta0}^{\gamma}\right]  _{\mathcal{L}}\right\}  Y_{\gamma
}^{m}\text{ }, \label{A16}%
\end{equation}%
\begin{equation}
Z_{\alpha\beta}=\frac{1}{2}\left[  \Gamma_{\alpha\beta}^{0}\right]
_{\mathcal{L}}\Lambda^{m}+\frac{1}{2}\left[  \Gamma_{\alpha\beta}^{\gamma
}\right]  _{\mathcal{L}}Y_{\gamma}^{m}\text{ }, \label{A17}%
\end{equation}%
\begin{equation}
\frac{dX^{m}}{dx^{0}}=\Lambda^{m}+Y_{\alpha}^{m}\frac{df^{\alpha}}{dx^{0}%
}\text{ }. \label{A18}%
\end{equation}
Because all $\Gamma$-symbols of the original metric and functions $f^{\alpha}$
are given equations (\ref{A15}) and (\ref{A16}) represent the closed linear
system of the ordinary differential equations of first order with respect to
the variable $x^{0}$ for coefficients $\Lambda^{m}(x^{0})$ and $Y_{\alpha}%
^{m}(x^{0})$ in expansion (\ref{A5}). These solutions should be substituted to
the right hand side of the equation (\ref{A17}) which gives coefficients
$Z_{\alpha\beta}(x^{0})$. After that we need to substitute $\Lambda^{m}%
(x^{0})$ and $Y_{\alpha}^{m}(x^{0})$ into the equation (\ref{A18}) where from
we obtain the last coefficients $X^{m}(x^{0})$ by quadrature.

This is the general procedure how to construct the Fermi coordinates for any
metric in vicinity of any given curve. There is also a possibility to
specialize the Fermi coordinates in such a way that the metric in these
coordinates in the first two approximations will be Minkowskian:%
\begin{equation}
\acute{g}_{ik}(\acute{x})=\eta_{ik}+O(2)\text{ }, \label{A19}%
\end{equation}
where $\eta_{ik}$ is Minkowski metric tensor. This can be done by choosing in
special way the arbitrary constants of integration which contain the general
solution of the equations (\ref{A15})-(\ref{A16}) and (\ref{A18}) (there are
20 such constants 10 of which should be fixed in order to obtain the form
(\ref{A19}) and another 10 will remain arbitrary reflecting the Poincar\`{e}
symmetry of the Minkowskian spacetime).

\section{Metric in Fermi coordinates}

The same line (\ref{A1}) in Fermi coordinates $\acute{x}$ has equation of the
similar form:%
\begin{equation}
\acute{x}^{\alpha}=F^{\alpha}\left(  \acute{x}^{0}\right)  \text{ }.
\label{A20}%
\end{equation}
The functions $F^{\alpha}\left(  \acute{x}^{0}\right)  $ follow from
transformation (\ref{A5}). This transformation tells that on the line
$\acute{x}^{0}=X^{0}(x^{0})$ and $\acute{x}^{\alpha}=X^{\alpha}(x^{0})$. Then
\begin{equation}
F^{\alpha}\left(  \acute{x}^{0}\right)  =[X^{\alpha}\left(  \zeta\right)
]_{\zeta=(arcX^{0})\left(  \acute{x}^{0}\right)  }\text{ }, \label{A21}%
\end{equation}
where $arcX^{0}$ is function inverse to $X^{0}$.

Because in Fermi coordinates
\begin{equation}
\left[  \acute{g}_{ik}\left(  \acute{x}\right)  \right]  _{\mathcal{L}}%
=c_{ik}\text{ },\text{ }\left[  \frac{\partial\acute{g}_{ik}\left(  \acute
{x}\right)  }{\partial\acute{x}^{l}}\right]  _{\mathcal{L}}=0\text{ },\text{
}c_{ik}=const \label{A22}%
\end{equation}
the expansion for metric near the line has the form:%
\begin{equation}
\acute{g}_{ik}\left(  \acute{x}\right)  =c_{ik}+\frac{1}{2}\left[
\frac{\partial^{2}\acute{g}_{ik}\left(  \acute{x}\right)  }{\partial\acute
{x}^{\alpha}\partial\acute{x}^{\beta}}\right]  _{\mathcal{L}}\left[  \acute
{x}^{\alpha}-F^{\alpha}(\acute{x}^{0})\right]  \left[  \acute{x}^{\beta
}-F^{\beta}(\acute{x}^{0})\right]  +O\left(  3\right)  \text{ }. \label{A23}%
\end{equation}
Then to obtain this metric we need the second derivatives of the metric tensor
with respect to the space coordinates $\acute{x}^{\alpha}$ on the line.
However, these second derivatives depend on the cubic terms $O\left(
3\right)  $ in expansion (\ref{A5}) and up to now remain completely arbitrary.
To make a choice for this cubic addend it is necessary to accept some
additional coordinate restrictions which will not violate conditions
(\ref{A22}). We already mentioned in Introduction that the natural physical
arguments for such a choice have been proposed by A. Eddington and here we
will follow his proposal, that is we will specify the cubic addends in
coordinates transformation to the Fermi coordinates so as to represent the
second derivatives in metric (\ref{A23}) in terms of the Riemann tensor.
Eddington showed that Riemann coordinates can be further specified in such a
way that cyclic combination $\acute{\Gamma}_{kl,m}^{i}+\acute{\Gamma}%
_{mk,l}^{i}+\acute{\Gamma}_{lm,k}^{i}$ of derivatives of $\acute{\Gamma}%
$-symbols at point where $\acute{\Gamma}_{kl}^{i}$ are zero also vanish. Under
this condition it is simple matter to express second derivatives of the metric
at this point in terms of the components of the Riemann tensor. In case of
Fermi coordinates we described in the preceding section the full 4-dimensional
Eddington condition cannot be accepted because it contradicts to the equations
(\ref{A15})-(\ref{A18}). However, it is possible to restrict the choice of
Fermi coordinates by the following reduced version of the same condition:
\begin{equation}
\left[  \frac{\partial\acute{\Gamma}_{\nu\lambda}^{i}\left(  \acute{x}\right)
}{\partial\acute{x}^{\mu}}+\frac{\partial\acute{\Gamma}_{\mu\nu}^{i}\left(
\acute{x}\right)  }{\partial\acute{x}^{\lambda}}+\frac{\partial\acute{\Gamma
}_{\lambda\mu}^{i}\left(  \acute{x}\right)  }{\partial\acute{x}^{\nu}}\right]
_{\mathcal{L}}=0\text{ }, \label{A25}%
\end{equation}
where the upper index remains 4-dimensional and all three lower indices are
3-dimensional. The proof of the possibility of this restriction we placed in
Appendix B.

Under the restriction $\left[  \acute{\Gamma}_{kl}^{i}\left(  \acute
{x}\right)  \right]  _{\mathcal{L}}=0$ from the general expression for the
Riemann tensor we have:%
\begin{equation}
\left[  \acute{R}_{klm}^{i}\left(  \acute{x}\right)  \right]  _{\mathcal{L}%
}=\left[  \frac{\partial\acute{\Gamma}_{km}^{i}\left(  \acute{x}\right)
}{\partial\acute{x}^{l}}-\frac{\partial\acute{\Gamma}_{kl}^{i}\left(
\acute{x}\right)  }{\partial\acute{x}^{m}}\right]  _{\mathcal{L}}\text{ }.
\label{A25-1}%
\end{equation}
Let's apply this formula for the 3-dimensional indices $(k,l,m)=(\nu
,\lambda,\mu)$ that is:
\begin{equation}
\left[  \acute{R}_{\nu\lambda\mu}^{i}\left(  \acute{x}\right)  \right]
_{\mathcal{L}}=\left[  \frac{\partial\acute{\Gamma}_{\nu\mu}^{i}\left(
\acute{x}\right)  }{\partial\acute{x}^{\lambda}}-\frac{\partial\acute{\Gamma
}_{\nu\lambda}^{i}\left(  \acute{x}\right)  }{\partial\acute{x}^{\mu}}\right]
_{\mathcal{L}}\text{ }. \label{A26}%
\end{equation}
By simple manipulation with indices it is easy to show that the last
expression with the help of condition (\ref{A25}) can be inverted:
\begin{equation}
\left[  \frac{\partial\acute{\Gamma}_{\nu\lambda}^{i}\left(  \acute{x}\right)
}{\partial\acute{x}^{\mu}}\right]  _{\mathcal{L}}=-\frac{1}{3}\left[
\acute{R}_{\nu\lambda\mu}^{i}\left(  \acute{x}\right)  +\acute{R}_{\lambda
\nu\mu}^{i}\left(  \acute{x}\right)  \right]  _{\mathcal{L}}\text{ }.
\label{A27}%
\end{equation}

Now from the identity $\left[  \acute{g}_{ik}\left(  \acute{x}\right)
\right]  _{;l;m}=0$, taking into account the restriction $\left[
\acute{\Gamma}_{kl}^{i}\left(  \acute{x}\right)  \right]  _{\mathcal{L}}=0$,
one can express the second derivatives of the metric tensor on the line
$\mathcal{L}$ in Fermi coordinates in the form:%
\begin{equation}
\left[  \frac{\partial\acute{g}_{ik}\left(  \acute{x}\right)  }{\partial
\acute{x}^{\lambda}\partial\acute{x}^{\mu}}\right]  _{\mathcal{L}}=\left[
\frac{\partial\acute{\Gamma}_{i\lambda}^{l}\left(  \acute{x}\right)
}{\partial\acute{x}^{\mu}}\acute{g}_{lk}\left(  \acute{x}\right)
+\frac{\partial\acute{\Gamma}_{k\lambda}^{l}\left(  \acute{x}\right)
}{\partial\acute{x}^{\mu}}\acute{g}_{li}\left(  \acute{x}\right)  \right]
_{\mathcal{L}}\text{ }. \label{A28}%
\end{equation}
From this formula we have:%
\begin{equation}
\left[  \frac{\partial\acute{g}_{00}\left(  \acute{x}\right)  }{\partial
\acute{x}^{\lambda}\partial\acute{x}^{\mu}}\right]  _{\mathcal{L}}=2\left[
\frac{\partial\acute{\Gamma}_{0\lambda}^{l}\left(  \acute{x}\right)
}{\partial\acute{x}^{\mu}}\acute{g}_{l0}\left(  \acute{x}\right)  \right]
_{\mathcal{L}}\text{ }, \label{A29}%
\end{equation}%
\begin{equation}
\left[  \frac{\partial\acute{g}_{0\alpha}\left(  \acute{x}\right)  }%
{\partial\acute{x}^{\lambda}\partial\acute{x}^{\mu}}\right]  _{\mathcal{L}%
}=\left[  \frac{\partial\acute{\Gamma}_{0\lambda}^{l}\left(  \acute{x}\right)
}{\partial\acute{x}^{\mu}}\acute{g}_{l\alpha}\left(  \acute{x}\right)
+\frac{\partial\acute{\Gamma}_{\alpha\lambda}^{l}\left(  \acute{x}\right)
}{\partial\acute{x}^{\mu}}\acute{g}_{l0}\left(  \acute{x}\right)  \right]
_{\mathcal{L}}\text{ }, \label{A30}%
\end{equation}%
\begin{equation}
\left[  \frac{\partial\acute{g}_{\alpha\beta}\left(  \acute{x}\right)
}{\partial\acute{x}^{\lambda}\partial\acute{x}^{\mu}}\right]  _{\mathcal{L}%
}=\left[  \frac{\partial\acute{\Gamma}_{\alpha\lambda}^{l}\left(  \acute
{x}\right)  }{\partial\acute{x}^{\mu}}\acute{g}_{l\beta}\left(  \acute
{x}\right)  +\frac{\partial\acute{\Gamma}_{\beta\lambda}^{l}\left(  \acute
{x}\right)  }{\partial\acute{x}^{\mu}}\acute{g}_{l\alpha}\left(  \acute
{x}\right)  \right]  _{\mathcal{L}}\text{ }. \label{A31}%
\end{equation}
The first two of these formulas show that in order to express all second
derivatives of the metric in terms of the Riemann tensor the relation
(\ref{A27}) is not enough. It is necessary to find analogous expression also
for the quantity $\partial\acute{\Gamma}_{0\lambda}^{l}\left(  \acute
{x}\right)  /\partial\acute{x}^{\mu}$ on the line. To do this let's take the
general 4-dimensional relation (\ref{A25-1}) for indices $k=\nu,l=\lambda,m=0$
and sum it with equation (\ref{A26}) being multiplied by the derivative
$dF^{\mu}\left(  \acute{x}^{0}\right)  /d\acute{x}^{0}$. In the right hand
side of this sum will appear the quantity%
\begin{equation}
\left[  \frac{\partial\acute{\Gamma}_{\nu\lambda}^{i}\left(  \acute{x}\right)
}{\partial\acute{x}^{0}}\right]  _{\mathcal{L}}+\left[  \frac{\partial
\acute{\Gamma}_{\nu\lambda}^{i}\left(  \acute{x}\right)  }{\partial\acute
{x}^{\mu}}\right]  _{\mathcal{L}}\frac{dF^{\mu}\left(  \acute{x}^{0}\right)
}{d\acute{x}^{0}}\text{ }, \label{A32}%
\end{equation}
which is zero because for any function $\acute{\Psi}\left(  \acute{x}\right)
$ which is zero along line $\mathcal{L}$, that is which satisfy the
restriction $\acute{\Psi}\left[  \acute{x}^{0},F^{1}\left(  \acute{x}%
^{0}\right)  ,F^{2}\left(  \acute{x}^{0}\right)  ,F^{3}\left(  \acute{x}%
^{0}\right)  \right]  =0$, the ordinary derivative of its value on the line
with respect to $\acute{x}^{0}$ is also zero and due this evident fact we
deduce:%
\begin{equation}
\frac{d}{d\acute{x}^{0}}\acute{\Psi}\left[  \acute{x}^{0},F^{1}\left(
\acute{x}^{0}\right)  ,F^{2}\left(  \acute{x}^{0}\right)  ,F^{3}\left(
\acute{x}^{0}\right)  \right]  =\left[  \frac{\partial\acute{\Psi}\left(
\acute{x}\right)  }{\partial\acute{x}^{0}}\right]  _{\mathcal{L}}+\left[
\frac{\partial\acute{\Psi}\left(  \acute{x}\right)  }{\partial\acute{x}^{\mu}%
}\right]  _{\mathcal{L}}\frac{dF^{\mu}\left(  \acute{x}^{0}\right)  }%
{d\acute{x}^{0}}=0\text{ }. \label{A33}%
\end{equation}
Then the resulting sum gives the following equation:%

\begin{equation}
\left[  \acute{R}_{\nu\lambda0}^{i}\left(  \acute{x}\right)  \right]
_{\mathcal{L}}+\left[  \acute{R}_{\nu\lambda\mu}^{i}\left(  \acute{x}\right)
\right]  _{\mathcal{L}}\frac{dF^{\mu}\left(  \acute{x}^{0}\right)  }%
{d\acute{x}^{0}}=\left[  \frac{\partial\acute{\Gamma}_{\nu0}^{i}\left(
\acute{x}\right)  }{\partial\acute{x}^{\lambda}}\right]  _{\mathcal{L}%
}+\left[  \frac{\partial\acute{\Gamma}_{\nu\mu}^{i}\left(  \acute{x}\right)
}{\partial\acute{x}^{\lambda}}\right]  _{\mathcal{L}}\frac{dF^{\mu}\left(
\acute{x}^{0}\right)  }{d\acute{x}^{0}}\text{ }, \label{A34}%
\end{equation}
where from the quantity $[\partial\acute{\Gamma}_{\nu0}^{i}\left(  \acute
{x}\right)  /\partial\acute{x}^{\lambda}]_{\mathcal{L}}$ can be represented in
terms of the Riemann tensor since for the derivatives $\left[  \partial
\acute{\Gamma}_{\nu\mu}^{i}\left(  \acute{x}\right)  /\partial\acute
{x}^{\lambda}\right]  _{\mathcal{L}}$\ we already have such representation,
see formula (\ref{A27}). The result is:%
\begin{equation}
\left[  \frac{\partial\acute{\Gamma}_{\nu0}^{i}\left(  \acute{x}\right)
}{\partial\acute{x}^{\lambda}}\right]  _{\mathcal{L}}=\left[  \acute{R}%
_{\nu\lambda0}^{i}\left(  \acute{x}\right)  \right]  _{\mathcal{L}}+\frac
{1}{3}[\acute{R}_{\mu\nu\lambda}^{i}\left(  \acute{x}\right)  -2\acute{R}%
_{\nu\mu\lambda}^{i}]_{\mathcal{L}}\frac{dF^{\mu}\left(  \acute{x}^{0}\right)
}{d\acute{x}^{0}}\text{ }. \label{A35}%
\end{equation}

Now from (\ref{A23}) and (\ref{A29})-(\ref{A31}) (using definition
$R_{iklm}=g_{in}R_{klm}^{n}$) we obtain the final general \cite{SS3} result
for the canonical (Eddington's terminology) metric in Fermi coordinates:%
\begin{gather}
\acute{g}_{00}\left(  \acute{x}\right)  =c_{00}\label{A36}\\
+\left[  \acute{R}_{0\lambda\mu0}\left(  \acute{x}\right)  -\frac{2}{3}%
\acute{R}_{0\lambda\nu\mu}\left(  \acute{x}\right)  \frac{dF^{\nu}\left(
\acute{x}^{0}\right)  }{d\acute{x}^{0}}\right]  _{\mathcal{L}}\left[
\acute{x}^{\lambda}-F^{\lambda}(\acute{x}^{0})\right]  \left[  \acute{x}^{\mu
}-F^{\mu}(\acute{x}^{0})\right]  +O\left(  3\right)  \text{ },\nonumber
\end{gather}%
\begin{gather}
\acute{g}_{0\alpha}\left(  \acute{x}\right)  =c_{0\alpha}\label{A37}\\
+\left[  \frac{2}{3}\acute{R}_{\alpha\lambda\mu0}\left(  \acute{x}\right)
-\frac{1}{3}\acute{R}_{\alpha\lambda\nu\mu}\left(  \acute{x}\right)
\frac{dF^{\nu}\left(  \acute{x}^{0}\right)  }{d\acute{x}^{0}}\right]
_{\mathcal{L}}\left[  \acute{x}^{\lambda}-F^{\lambda}(\acute{x}^{0})\right]
\left[  \acute{x}^{\mu}-F^{\mu}(\acute{x}^{0})\right]  +O\left(  3\right)
\text{ },\nonumber
\end{gather}

\begin{equation}
\acute{g}_{\alpha\beta}\left(  \acute{x}\right)  =c_{\alpha\beta}+\frac{1}%
{3}\left[  \acute{R}_{\alpha\lambda\mu\beta}\left(  \acute{x}\right)  \right]
_{\mathcal{L}}\left[  \acute{x}^{\lambda}-F^{\lambda}(\acute{x}^{0})\right]
\left[  \acute{x}^{\mu}-F^{\mu}(\acute{x}^{0})\right]  +O\left(  3\right)
\text{ }. \label{A38}%
\end{equation}

\section{Fermi coordinates for static observer in Schwarzschild spacetime}

Let's take the Schwarzschild metric in its standard form:%
\begin{equation}
-ds^{2}=-\left(  1-\frac{2m}{r}\right)  dt^{2}+\left(  1-\frac{2m}{r}\right)
^{-1}dr^{2}+r^{2}\left(  d\theta^{2}+\sin^{2}\theta d\varphi^{2}\right)
\text{ }, \label{A39}%
\end{equation}
with following designation for coordinates:%
\begin{equation}
t,r,\theta,\varphi=x^{0},x^{1},x^{2},x^{3}. \label{A40}%
\end{equation}
The world line of a static observer is:%
\begin{equation}
x^{\alpha}=x_{\ast}^{\alpha} \label{A41}%
\end{equation}
where $x_{\ast}^{\alpha}=(x_{\ast}^{1},x_{\ast}^{2},x_{\ast}^{3})=(r_{\ast
},\theta_{\ast},\varphi_{\ast})$ are arbitrary constants. The transformation
to Fermi coordinates $\acute{x}$ along this line is given by the formula
(\ref{A5}), that is%
\begin{equation}
\acute{x}^{m}=X^{m}(t)+Y_{\alpha}^{m}(t)(x^{\alpha}-x_{\ast}^{\alpha
})+Z_{\alpha\beta}^{m}(t)(x^{\alpha}-x_{\ast}^{\alpha})(x^{\beta}-x_{\ast
}^{\beta})+O(3)\text{ }. \label{A42}%
\end{equation}
In equations (\ref{A15})-(\ref{A16}) and (\ref{A18}) all terms containing
$df^{\alpha}/dx^{0}$ disappear and among those $\Gamma-$symbols which are
present in these equations there are only two non-zero, namely%
\begin{equation}
\lbrack\Gamma_{00}^{1}]_{\mathcal{L}}=\frac{m}{r_{\ast}^{2}}\left(
1-\frac{2m}{r_{\ast}}\right)  \text{ },\text{ }[\Gamma_{10}^{0}]_{\mathcal{L}%
}=\frac{m}{r_{\ast}^{2}}\left(  1-\frac{2m}{r_{\ast}}\right)  ^{-1}.
\label{A43}%
\end{equation}
Under these conditions equations (\ref{A15})-(\ref{A16}) and (\ref{A18})
become very simple and can be integrated easily. The solution for the
functions $\Lambda^{m}\left(  t\right)  $ is $\Lambda^{m}=C_{1}^{m}e^{\omega
t}+C_{2}^{m}e^{-\omega t}$ and for coefficients $X^{m}\left(  t\right)  $ and
$Y_{\alpha}^{m}(t)$ we have:%
\begin{equation}
X^{m}=\omega^{-1}\left(  C_{1}^{m}e^{\omega t}-C_{2}^{m}e^{-\omega t}\right)
+C_{3}^{m}\text{ }, \label{A44}%
\end{equation}%
\begin{equation}
Y_{1}^{m}=\left(  1-\frac{2m}{r_{\ast}}\right)  ^{-1}\left(  C_{1}%
^{m}e^{\omega t}-C_{2}^{m}e^{-\omega t}\right)  \text{ }, \label{A45}%
\end{equation}%
\begin{equation}
Y_{2}^{m}=C_{4}^{m}\text{ },\text{ }Y_{3}^{m}=C_{5}^{m}\text{ }, \label{A46}%
\end{equation}
where $C_{1}^{m},...,C_{5}^{m}$ are arbitrary constants of integration and%
\begin{equation}
\omega=\frac{m}{r_{\ast}^{2}}\text{ }. \label{A47}%
\end{equation}
Without loss of generality we can chose constants $C_{1}^{m},C_{2}^{m}%
,C_{4}^{m},C_{5}^{m}$ in the following way:%
\begin{equation}
C_{1}^{m}=\left(  C_{1}^{0},C_{1}^{1},C_{1}^{2},C_{1}^{3}\right)  =\left(
\lambda,\lambda,0,0\right)  \text{ }, \label{A48}%
\end{equation}%
\begin{equation}
C_{2}^{m}=\left(  C_{2}^{0},C_{2}^{1},C_{2}^{2},C_{2}^{3}\right)  =\left(
\lambda,-\lambda,0,0\right)  \text{ }, \label{A49}%
\end{equation}%
\begin{equation}
C_{4}^{m}=\left(  C_{4}^{0},C_{4}^{1},C_{4}^{2},C_{4}^{3}\right)  =\left(
0,0,r_{\ast},0\right)  \text{ }, \label{A50}%
\end{equation}%
\begin{equation}
C_{5}^{m}=\left(  C_{5}^{0},C_{5}^{1},C_{5}^{2},C_{5}^{3}\right)  =\left(
0,0,0,r_{\ast}\sin\theta_{\ast}\right)  \text{ }, \label{A51}%
\end{equation}
where quantity $\lambda$ is defined by the relation%
\begin{equation}
\lambda^{2}=\frac{1}{4}\left(  1-\frac{2m}{r_{\ast}}\right)  \text{ }.
\label{A52}%
\end{equation}
This choice for free parameters fixes the arbitrary constants $c_{ik}$ in the
metric (\ref{A36})-(\ref{A38}) as
\begin{equation}
c_{00}=-1,\text{ }c_{0\alpha}=0,\text{ }c_{\alpha\beta}=\delta_{\alpha\beta
}\text{ }, \label{A53}%
\end{equation}
that is in the first two approximations the metric is Minkowskian in Fermi coordinates.

Now we substitute the constants (\ref{A48})-(\ref{A51}) into expressions
(\ref{A44})-(\ref{A46}) to obtain the final form for coefficients $X^{m}(t)$,
$Y_{\alpha}^{m}(t)$ and after that insert them together with Schwarzschild
$\Gamma$-symbols $\left[  \Gamma_{\alpha\beta}^{0}\right]  _{\mathcal{L}}$ and
$\left[  \Gamma_{\alpha\beta}^{\gamma}\right]  _{\mathcal{L}}$ into the right
hand side of the equation (\ref{A17}). This gives the coefficients
$Z_{\alpha\beta}\left(  t\right)  $ after which we can write the final form of
transformation to Fermi coordinates along the world line of static
Schwarzschild observer:%
\begin{gather}
\acute{x}^{0}=C_{3}^{0}+\frac{2\lambda}{\omega}\sinh\omega t+\frac{1}%
{2\lambda}\left(  r-r_{\ast}\right)  \sinh\omega t\label{A54}\\
-\left[  \frac{\omega}{16\lambda^{3}}\left(  r-r_{\ast}\right)  ^{2}+r_{\ast
}\lambda\left(  \theta-\theta_{\ast}\right)  ^{2}+r_{\ast}\lambda\sin
^{2}\theta_{\ast}\left(  \varphi-\varphi_{\ast}\right)  ^{2}\right]
\sinh\omega t+O\left(  3\right)  \text{ },\nonumber
\end{gather}%
\begin{gather}
\acute{x}^{1}=C_{3}^{1}+\frac{2\lambda}{\omega}\cosh\omega t+\frac{1}%
{2\lambda}\left(  r-r_{\ast}\right)  \cosh\omega t\label{A55}\\
-\left[  \frac{\omega}{16\lambda^{3}}\left(  r-r_{\ast}\right)  ^{2}+r_{\ast
}\lambda\left(  \theta-\theta_{\ast}\right)  ^{2}+r_{\ast}\lambda\sin
^{2}\theta_{\ast}\left(  \varphi-\varphi_{\ast}\right)  ^{2}\right]
\cosh\omega t+O\left(  3\right)  \text{ },\nonumber
\end{gather}%
\begin{equation}
\acute{x}^{2}=C_{3}^{2}+r_{\ast}\left(  \theta-\theta_{\ast}\right)  +\left(
r-r_{\ast}\right)  \left(  \theta-\theta_{\ast}\right)  -\frac{1}{2}r_{\ast
}\sin\theta_{\ast}\cos\theta_{\ast}\left(  \varphi-\varphi_{\ast}\right)
^{2}+O\left(  3\right)  \text{ }, \label{A56}%
\end{equation}%
\begin{gather}
\acute{x}^{3}=C_{3}^{3}+r_{\ast}\sin\theta_{\ast}\left(  \varphi-\varphi
_{\ast}\right)  +\sin\theta_{\ast}\left(  r-r_{\ast}\right)  \left(
\varphi-\varphi_{\ast}\right) \label{A57}\\
+r_{\ast}\cos\theta_{\ast}\left(  \theta-\theta_{\ast}\right)  \left(
\varphi-\varphi_{\ast}\right)  +O\left(  3\right)  \text{ }.\nonumber
\end{gather}

Metric for the static Schwarzschild observer in canonical Fermi coordinates
follows from formulas (\ref{A36})-(\ref{A38}). The arbitrary constants
$c_{ik}$ we already specified, see (\ref{A53}). Now we need to find the
functions $F^{\alpha}\left(  \acute{x}^{0}\right)  $ and components of the
Riemann tensor $\acute{R}_{iklm}\left(  \acute{x}\right)  $. The equation of
the Schwarzschild static world line in the Fermi coordinates can be extracted
from transformation (\ref{A54})-(\ref{A57}). On the line we have%
\begin{equation}
\acute{x}^{0}=C_{3}^{0}+\frac{2\lambda}{\omega}\sinh\omega t\text{ },\text{
}\acute{x}^{1}=C_{3}^{1}+\frac{2\lambda}{\omega}\cosh\omega t\text{ },\text{
}\acute{x}^{2}=C_{3}^{2}\text{ },\text{ }\acute{x}^{3}=C_{3}^{3}\text{ }.
\label{A58}%
\end{equation}
Then functions $F^{\alpha}\left(  \acute{x}^{0}\right)  $ are:%
\begin{equation}
F^{1}\left(  \acute{x}^{0}\right)  =C_{3}^{1}+\sqrt{a+\left(  \acute{x}%
^{0}-C_{3}^{0}\right)  ^{2}}\text{ },\text{ }F^{2}=C_{3}^{2}\text{ },\text{
}F^{3}=C_{3}^{3}\text{ }, \label{A59}%
\end{equation}
where
\begin{equation}
a=\frac{r_{\ast}^{4}}{m^{2}}\left(  1-\frac{2m}{r_{\ast}}\right)  \text{ }.
\label{A59-1}%
\end{equation}
The arbitrary constants $C_{3}^{i}$ are not important, they can be eliminated
by the shift of the origin of the Fermi coordinates.

The Riemann tensor $\acute{R}_{iklm}\left(  \acute{x}\right)  $ can be found
by transformation (\ref{A54})-(\ref{A57}) from its known counterpart
$R_{iklm}\left(  x\right)  $ for the Schwarzschild metric (\ref{A39}) which
has the following non-zero components:%
\begin{equation}
R_{0101}=R_{trtr}=-\frac{2m}{r^{3}}\text{ }, \label{A60}%
\end{equation}%
\begin{equation}
R_{0202}=R_{t\theta t\theta}=\frac{m}{r}\left(  1-\frac{2m}{r}\right)  \text{
}, \label{A61}%
\end{equation}%
\begin{equation}
R_{0303}=R_{t\varphi t\varphi}=\frac{m}{r}\left(  1-\frac{2m}{r}\right)
\sin^{2}\theta\text{ }, \label{A62}%
\end{equation}%
\begin{equation}
R_{1212}=R_{r\theta r\theta}=-\frac{m}{r}\left(  1-\frac{2m}{r}\right)
^{-1}\text{ }, \label{A63}%
\end{equation}%
\begin{equation}
R_{1313}=R_{r\varphi r\varphi}=-\frac{m}{r}\left(  1-\frac{2m}{r}\right)
^{-1}\sin^{2}\theta\text{ }, \label{A64}%
\end{equation}%
\begin{equation}
R_{2323}=R_{\theta\varphi\theta\varphi}=2mr\sin^{2}\theta\text{ }. \label{A65}%
\end{equation}
We do not included in this list those non-zero components of $R_{iklm}\left(
x\right)  $ which can be obtained from (\ref{A60})-(\ref{A65}) by application
of all symmetries of the Riemann tensor. These components transform to the
components of $\acute{R}_{iklm}\left(  \acute{x}\right)  $ by the usual tensor
law and on the line this transformation takes the form:%
\begin{equation}
\lbrack\acute{R}_{psqn}]_{\mathcal{L}}=[R_{iklm}Q_{p}^{i}Q_{s}^{k}Q_{q}%
^{l}Q_{n}^{m}]_{\mathcal{L}}\text{ }, \label{A66}%
\end{equation}
where matrix $Q_{k}^{i}$ is inverse to the Jacobian matrix $A_{k}^{i}$
introduced in (\ref{A2}), see also (\ref{P9}). For the transformation
(\ref{A54})-(\ref{A57}) these matrices calculated on the line $\mathcal{L}$
(the upper index numerates the matrix lines and lower index corresponds to the
columns) are:%
\begin{equation}
\left[  A_{k}^{i}\right]  _{\mathcal{L}}=\left(
\begin{array}
[c]{cccc}%
2\lambda\cosh\omega t & \left(  2\lambda\right)  ^{-1}\sinh\omega t & 0 & 0\\
2\lambda\sinh\omega t & \left(  2\lambda\right)  ^{-1}\cosh\omega t & 0 & 0\\
0 & 0 & r_{\ast} & 0\\
0 & 0 & 0 & r_{\ast}\sin\theta_{\ast}%
\end{array}
\right)  \text{ }, \label{A67}%
\end{equation}%
\begin{equation}
\left[  Q_{k}^{i}\right]  _{\mathcal{L}}=\left(
\begin{array}
[c]{cccc}%
(2\lambda)^{-1}\cosh\omega t & -\left(  2\lambda\right)  ^{-1}\sinh\omega t &
0 & 0\\
-2\lambda\sinh\omega t & 2\lambda\cosh\omega t & 0 & 0\\
0 & 0 & (r_{\ast})^{-1} & 0\\
0 & 0 & 0 & (r_{\ast}\sin\theta_{\ast})^{-1}%
\end{array}
\right)  \text{ }. \label{A68}%
\end{equation}
Calculations of $[\acute{R}_{psqn}]_{\mathcal{L}}$ from (\ref{A66}) using
$\left[  Q_{k}^{i}\right]  _{\mathcal{L}}$ from (\ref{A68}) and $[R_{iklm}%
]_{\mathcal{L}}=$ $R_{iklm}\left(  r_{\ast},\theta_{\ast}\right)  $ from
(\ref{A60})-(\ref{A65}) give:%

\begin{equation}
\left[  \acute{R}_{0101}\right]  _{\mathcal{L}}=-\frac{2m}{r_{\ast}^{3}%
},\text{ }\left[  \acute{R}_{0202}\right]  _{\mathcal{L}}=\frac{m}{r_{\ast
}^{3}},\text{ }\left[  \acute{R}_{0303}\right]  _{\mathcal{L}}=\frac
{m}{r_{\ast}^{3}}, \label{A69}%
\end{equation}%
\begin{equation}
\left[  \acute{R}_{1212}\right]  _{\mathcal{L}}=-\frac{m}{r_{\ast}^{3}},\text{
}\left[  \acute{R}_{1313}\right]  _{\mathcal{L}}=-\frac{m}{r_{\ast}^{3}%
},\text{ }\left[  \acute{R}_{2323}\right]  _{\mathcal{L}}=\frac{2m}{r_{\ast
}^{3}}\text{ }. \label{A70}%
\end{equation}
We see that on line $\mathcal{L}$ the Riemann tensor in the Fermi coordinates
contains the same set of non-zero components as in Schwarzschild coordinates
but their values are simpler. We again do not included in formulas
(\ref{A69})-(\ref{A70}) those non-zero components of $\left[  \acute{R}%
_{iklm}\right]  _{\mathcal{L}}$ which can be obtained by application the
symmetries of the Riemann tensor.

To write down the final form of the metric it is convenient to introduce
shifting Fermi coordinates $\tau,u,v,w$:%
\begin{equation}
\tau=\acute{x}^{0}-C_{3}^{0}\text{ },\text{ }u=\acute{x}^{1}-C_{3}^{1}\text{
},\text{ }v=\acute{x}^{2}-C_{3}^{2}\text{ },\text{ }w=\acute{x}^{3}-C_{3}%
^{3}\text{ }. \label{A71}%
\end{equation}
Collecting all information on the constants $c_{ik}$ (\ref{A53}), functions
$F^{\alpha}\left(  \acute{x}^{0}\right)  $ (\ref{A59}), and components of the
Riemann tensor $[\acute{R}_{iklm}\left(  \acute{x}\right)  ]_{\mathcal{L}}$
(\ref{A69})-(\ref{A70}) we obtain from (\ref{A36})-(\ref{A38}) the final form
of the metric for the static Schwarzschild observer in Fermi coordinates
$\tau,u,v,w$ (\ref{A71}):
\begin{gather}
-ds^{2}=\acute{g}_{ik}\left(  \acute{x}\right)  d\acute{x}^{i}d\acute{x}%
^{k}=\acute{g}_{\tau\tau}d\tau^{2}+2\acute{g}_{\tau u}d\tau du+2\acute
{g}_{\tau v}d\tau dv+2\acute{g}_{\tau w}d\tau dw\label{A72}\\
+\acute{g}_{uu}du^{2}+\acute{g}_{vv}dv^{2}+\acute{g}_{ww}dw^{2}+2\acute
{g}_{uv}dudv+2\acute{g}_{uw}dudw+2\acute{g}_{vw}dvdw\text{ },\nonumber
\end{gather}
where components of the metric tensor (up to the quadratic terms with respect
to the three small deviations $u-\sqrt{\tau^{2}+a}$ $,v,$ $w$ from the line)
are:%
\begin{equation}
\acute{g}_{\tau\tau}=-1+\frac{m}{r_{\ast}^{3}}\left[  2\left(  u-\sqrt
{\tau^{2}+a}\right)  ^{2}-v^{2}-w^{2}\right]  \text{ }, \label{A73}%
\end{equation}%
\begin{equation}
\acute{g}_{uu}=1+\frac{m}{3r_{\ast}^{3}}\left(  v^{2}+w^{2}\right)  \text{ },
\label{A74}%
\end{equation}%
\begin{equation}
\acute{g}_{vv}=1+\frac{m}{3r_{\ast}^{3}}\left[  \left(  u-\sqrt{\tau^{2}%
+a}\right)  ^{2}-2w^{2}\right]  \text{ }, \label{A75}%
\end{equation}%
\begin{equation}
\acute{g}_{ww}=1+\frac{m}{3r_{\ast}^{3}}\left[  \left(  u-\sqrt{\tau^{2}%
+a}\right)  ^{2}-2v^{2}\right]  \text{ }, \label{A76}%
\end{equation}

\begin{equation}
\acute{g}_{\tau u}=\frac{m\tau\left(  v^{2}+w^{2}\right)  }{3r_{\ast}^{3}%
\sqrt{\tau^{2}+a}},\text{ }\acute{g}_{\tau v}=\frac{m\tau\left(  \sqrt
{\tau^{2}+a}-u\right)  v}{3r_{\ast}^{3}\sqrt{\tau^{2}+a}},\text{ }\acute
{g}_{\tau w}=\frac{m\tau\left(  \sqrt{\tau^{2}+a}-u\right)  w}{3r_{\ast}%
^{3}\sqrt{\tau^{2}+a}}, \label{A77}%
\end{equation}%
\begin{equation}
\acute{g}_{uv}=\frac{m}{3r_{\ast}^{3}}\left(  \sqrt{\tau^{2}+a}-u\right)
v,\text{ }\acute{g}_{uw}=\frac{m}{3r_{\ast}^{3}}\left(  \sqrt{\tau^{2}%
+a}-u\right)  w,\text{ }\acute{g}_{vw}=\frac{2m}{3r_{\ast}^{3}}vw. \label{A78}%
\end{equation}

To understand better relation between Synge and Fermi approach it would be
instructive to take a timelike non-geodesic line and construct along it two
different coordinate systems: 1) Fermi coordinates and 2) Singe's quasi-Fermi
coordinates (that is coordinates associated with the observer's proper
reference frame along this line) and work out the transformation between these
two coordinate systems. In general this is not simple enterprise, however, in
particular case of the static observer in Schwarzschild spacetime considered
in this section the task can be resolved easily. In this case Fermi
coordinates along static world line in terms of Schwarzschild coordinates we
already found [see (\ref{A54})-(\ref{A57})]. For simplicity let's take in
these formulas $\theta_{\ast}=\pi/2$ in which case the Fermi coordinates
$\tau,u,v,w$ (\ref{A71}) are:
\begin{equation}
\tau=\rho\sinh\omega t+O\left(  3\right)  \text{ }, \label{A79}%
\end{equation}%
\begin{equation}
u=\rho\cosh\omega t+O\left(  3\right)  \text{ }, \label{A80}%
\end{equation}%
\begin{equation}
v=r_{\ast}\left(  \theta-\pi/2\right)  +\left(  r-r_{\ast}\right)  \left(
\theta-\pi/2\right)  +O\left(  3\right)  \text{ }, \label{A81}%
\end{equation}%
\begin{equation}
w=r_{\ast}\left(  \varphi-\varphi_{\ast}\right)  +\left(  r-r_{\ast}\right)
\left(  \varphi-\varphi_{\ast}\right)  +O\left(  3\right)  \text{ },
\label{A82}%
\end{equation}
where we introduced the notation:%
\begin{equation}
\rho=\frac{2\lambda}{\omega}+\frac{1}{2\lambda}\left(  r-r_{\ast}\right)
-\frac{\omega}{16\lambda^{3}}\left(  r-r_{\ast}\right)  ^{2}-r_{\ast}%
\lambda\left(  \theta-\pi/2\right)  ^{2}-r_{\ast}\lambda\left(  \varphi
-\varphi_{\ast}\right)  ^{2}. \label{A83}%
\end{equation}

The Synge's quasi-Fermi coordinates $T,X,Y_{,}Z$ \ along the same world line
in terms of the same Schwarzschild coordinates have been constructed in
\cite{BGJ} and they are \cite{SS4}:%
\begin{equation}
T=2\lambda t\text{ }, \label{A84}%
\end{equation}%
\begin{equation}
X=\rho-\frac{2\lambda}{\omega}+O\left(  3\right)  \label{A85}%
\end{equation}%
\begin{equation}
Y=r_{\ast}\left(  \theta-\pi/2\right)  +\left(  r-r_{\ast}\right)  \left(
\theta-\pi/2\right)  +O\left(  3\right)  \text{ }, \label{A86}%
\end{equation}%
\begin{equation}
Z=r_{\ast}\left(  \varphi-\varphi_{\ast}\right)  +\left(  r-r_{\ast}\right)
\left(  \varphi-\varphi_{\ast}\right)  +O\left(  3\right)  \text{ }.
\label{A87}%
\end{equation}
then from (\ref{A79})-(\ref{A87}) follows transformation between these two
coordinate systems:%
\begin{equation}
\tau=\left(  \frac{2\lambda}{\omega}+X\right)  \sinh\frac{\omega T}{2\lambda
}+O\left(  3\right)  \text{ }, \label{A88}%
\end{equation}%
\begin{equation}
u=\left(  \frac{2\lambda}{\omega}+X\right)  \cosh\frac{\omega T}{2\lambda
}+O\left(  3\right)  \text{ }, \label{A89}%
\end{equation}%
\begin{equation}
v=Y+O\left(  3\right)  \text{ },\text{ }w=Z+O\left(  3\right)  \text{ }.
\label{A90}%
\end{equation}
Because of spherical symmetry the angular coordinatization in both system
coincide as it should be. The transformation in the radial-time sector is of
Rindler-Minkowski type as also should be because (up to the second order with
respect to the deviation from the line) the proper frame of a static observer
in Schwarzschild metric (with Synge's coordinates $T,X$) is equivalent to
one-dimensional accelerated motion in flat space (with Fermi coordinates
$\tau,u$).

\section{Acknowledgment}

This paper is continuation of my work I started a year ago at International
Solvay Institutes (Brussels University). I would like to express my sincere
gratitude to Professor M. Henneaux for invitation and warm hospitality. I am
also grateful to anonymous referee whose comments and suggestions was very
useful for improvement the paper.

%

\appendix

\section{Standard formulas.}

We use notations of the the book \cite{LL}. In any spacetime with coordinates
$x^{i}$ and metric tensor $g_{ik}$ the $\Gamma$-symbols and Riemann tensor
are:%
\begin{equation}
\Gamma_{kl}^{i}=\frac{1}{2}g^{im}\left(  g_{mk,l}+g_{lm,k}-g_{kl,m}\right)
\text{ }, \label{P1}%
\end{equation}%
\begin{equation}
R_{klm}^{i}=\Gamma_{km,l}^{i}-\Gamma_{kl,m}^{i}+\Gamma_{nl}^{i}\Gamma_{km}%
^{n}-\Gamma_{nm}^{i}\Gamma_{kl}^{n}\text{ }, \label{P2}%
\end{equation}%
\begin{equation}
R_{iklm}=g_{in}R_{klm}^{n}\text{ }. \label{P3}%
\end{equation}
There are 4 symmetry identities for Riemann tensor:%
\begin{equation}
R_{iklm}=-R_{kilm}\text{ },\text{ }R_{iklm}=-R_{ikml}\text{ },\text{ }%
R_{iklm}=R_{lmik}\text{ }, \label{P4}%
\end{equation}%
\begin{equation}
R_{iklm}+R_{imkl}+R_{ilmk}=0\text{ }. \label{P5}%
\end{equation}
From definitions (\ref{P1})-(\ref{P3}) follows another representation for
$R_{iklm}$:%
\begin{equation}
R_{iklm}=\frac{1}{2}(g_{im,kl}+g_{kl,im}-g_{il,km}-g_{km,il})+g_{np}\left(
\Gamma_{kl}^{n}\Gamma_{im}^{p}-\Gamma_{km}^{n}\Gamma_{il}^{p}\right)  \text{
}. \label{P6}%
\end{equation}

\section{On the reduced Eddington coordinates restriction.}

The transformation (\ref{A5}) with cubic terms is:%

\begin{gather}
\acute{x}^{m}=X^{m}(x^{0})+Y_{\alpha}^{m}(x^{0})\left[  x^{\alpha}-f^{\alpha
}(x^{0})\right] \label{P7}\\
+Z_{\alpha\beta}^{m}(x^{0})\left[  x^{\alpha}-f^{\alpha}(x^{0})\right]
\left[  x^{\beta}-f^{\beta}(x^{0})\right] \nonumber\\
+W_{\alpha\beta\gamma}^{m}\left(  x^{0}\right)  \left[  x^{\alpha}-f^{\alpha
}(x^{0})\right]  \left[  x^{\beta}-f^{\beta}(x^{0})\right]  \left[  x^{\gamma
}-f^{\gamma}(x^{0})\right]  +O\left(  4\right)  \text{ },\nonumber
\end{gather}
where coefficients $W_{\alpha\beta\gamma}^{m}$ are symmetric with respect to
the transposition of any two of the lower indices. Then we have 40 (ten for
each 4-dimensional index $m$) independent coefficients $W_{\alpha\beta\gamma
}^{m}$. Now we apply the 4-dimensional partial derivative $\partial/\partial
x^{s}$ to the general transformation of $\Gamma$-symbols (\ref{A3}) and
restrict the result to the line $\mathcal{L}$ (taking into account that all
$\acute{\Gamma}_{nm}^{q}$ are zero on this line). This operation gives:%
\begin{equation}
\left[  \frac{\partial}{\partial x^{s}}\left(  \Gamma_{kl}^{i}A_{i}%
^{q}\right)  =\left(  \frac{\partial}{\partial\acute{x}^{p}}\acute{\Gamma
}_{nm}^{q}\right)  A_{s}^{p}A_{k}^{n}A_{l}^{m}+\frac{\partial^{3}\acute{x}%
^{q}}{\partial x^{k}\partial x^{l}\partial x^{s}}\right]  _{\mathcal{L}}\text{
}. \label{P8}%
\end{equation}
Let's denote the 4-dimensional matrix inverse to $A_{k}^{i}$ by $Q_{k}^{i}$,
that is:%
\begin{equation}
Q_{i}^{k}A_{k}^{l}=\delta_{i}^{l}\text{ }, \label{P9}%
\end{equation}
and multiply relation (\ref{P8}) by $(Q_{\alpha}^{s}Q_{\beta}^{k}Q_{\gamma
}^{l})_{\mathcal{L}}$ with all three lower indices 3-dimensional. We obtain:%
\begin{equation}
\left[  Q_{\alpha}^{s}Q_{\beta}^{k}Q_{\gamma}^{l}\frac{\partial}{\partial
x^{s}}\left(  \Gamma_{kl}^{i}A_{i}^{q}\right)  =\frac{\partial}{\partial
\acute{x}^{\alpha}}\acute{\Gamma}_{\beta\gamma}^{q}+\frac{\partial^{3}%
\acute{x}^{q}}{\partial x^{k}\partial x^{l}\partial x^{s}}Q_{\alpha}%
^{s}Q_{\beta}^{k}Q_{\gamma}^{l}\right]  _{\mathcal{L}}\text{ }. \label{P10}%
\end{equation}
Then we repeat this relation two times more with cyclic permutation of the
3-dimensional indices $\beta,\gamma,\alpha\rightarrow\alpha,\beta
,\gamma\rightarrow\gamma,\alpha,\beta$ and sum all three expressions. In
result we have:%
\begin{gather}
\left[  \frac{\partial}{\partial\acute{x}^{\alpha}}\acute{\Gamma}_{\beta
\gamma}^{q}+\frac{\partial}{\partial\acute{x}^{\gamma}}\acute{\Gamma}%
_{\alpha\beta}^{q}+\frac{\partial}{\partial\acute{x}^{\beta}}\acute{\Gamma
}_{\gamma\alpha}^{q}\right]  _{\mathcal{L}}=-\left\{  3Q_{\alpha}^{s}Q_{\beta
}^{k}Q_{\gamma}^{l}\frac{\partial^{3}\acute{x}^{q}}{\partial x^{k}\partial
x^{l}\partial x^{s}}\right\}  _{\mathcal{L}}\label{P11}\\
+\left\{  Q_{\alpha}^{s}Q_{\beta}^{k}Q_{\gamma}^{l}\left[  \frac{\partial
}{\partial x^{s}}\left(  \Gamma_{kl}^{i}A_{i}^{q}\right)  +\frac{\partial
}{\partial x^{l}}\left(  \Gamma_{sk}^{i}A_{i}^{q}\right)  +\frac{\partial
}{\partial x^{k}}\left(  \Gamma_{ls}^{i}A_{i}^{q}\right)  \right]  \right\}
_{\mathcal{L}}\text{ }.\nonumber
\end{gather}

Consequently the 3-dimensional Eddington condition (\ref{A25}) will be
satisfied if we chose the cubic addend in transformation (\ref{P7}) to satisfy
the requirement:%
\begin{gather}
\left\{  Q_{\alpha}^{s}Q_{\beta}^{k}Q_{\gamma}^{l}\frac{\partial^{3}\acute
{x}^{q}}{\partial x^{k}\partial x^{l}\partial x^{s}}\right\}  _{\mathcal{L}%
}\label{P12}\\
=\frac{1}{3}\left\{  Q_{\alpha}^{s}Q_{\beta}^{k}Q_{\gamma}^{l}\left[
\frac{\partial}{\partial x^{s}}\left(  \Gamma_{kl}^{i}A_{i}^{q}\right)
+\frac{\partial}{\partial x^{l}}\left(  \Gamma_{sk}^{i}A_{i}^{q}\right)
+\frac{\partial}{\partial x^{k}}\left(  \Gamma_{ls}^{i}A_{i}^{q}\right)
\right]  \right\}  _{\mathcal{L}}\text{ }.\nonumber
\end{gather}

The left and right sides in relation (\ref{P12}) are symmetric with respect to
the transposition of any two of the indices $\alpha,\beta,\gamma$,
consequently this relation represents 40 independent equations for 40 unknown
coefficients $W_{\alpha\beta\gamma}^{m}$ which enter the third derivatives of
$\acute{x}^{q}$. No other quantity in (\ref{P12}) contain these $W_{\alpha
\beta\gamma}^{m}$. It is important that terms $(\partial^{3}\acute{x}%
^{q}/\partial x^{k}\partial x^{l}\partial x^{s})_{\mathcal{L}}$ are linear
with respect to $W_{\alpha\beta\gamma}^{m}\left(  x^{0}\right)  $ and do not
contain $x^{0}$-derivatives of these functions. Then the system (\ref{P12}) is
the set of the linear algebraic equations with respect to the unknowns
$W_{\alpha\beta\gamma}^{m}$. \ Indeed, only the last term in expansion
(\ref{P7}) for Fermi coordinates $\acute{x}^{q}$ contains quantities
$W_{\alpha\beta\gamma}^{m}$ and it is easy to show that the left hand side of
equation (\ref{P12}) has the structure:
\begin{equation}
\left[  Q_{\alpha}^{s}Q_{\beta}^{k}Q_{\gamma}^{l}\frac{\partial^{3}\acute
{x}^{q}}{\partial x^{k}\partial x^{l}\partial x^{s}}\right]  _{\mathcal{L}%
}=6W_{\mu\lambda\nu}^{q}\left[  N_{\alpha}^{\mu}N_{\beta}^{\lambda}N_{\gamma
}^{\nu}\right]  _{\mathcal{L}}+...\text{ }, \label{P13}%
\end{equation}
where dots mean all terms which do not contain coefficients $W_{\mu\lambda\nu
}^{q}$ and $3\times3$ matrix $(N_{\beta}^{\alpha})_{\mathcal{L}}$ is:%
\begin{equation}
\left[  N_{\beta}^{\alpha}\right]  _{\mathcal{L}}=\left[  Q_{\beta}^{\alpha
}-Q_{\beta}^{0}\frac{df^{\alpha}}{dx^{0}}\right]  _{\mathcal{L}}\text{ }.
\label{P14}%
\end{equation}
Then using matrix inverse to $(N_{\beta}^{\alpha})_{\mathcal{L}}$ the system
(\ref{P12}) can be uniquely resolved with respect to the unknown coefficients
$W_{\mu\lambda\nu}^{q}$. This is the proof of the possibility to specialize
the Fermi coordinate in the way to achieve the 3-dimensional analogue of the
Eddington coordinates condition (\ref{A25}).

\begin{thebibliography}{99}
\bibitem {SS1}The Synge's quasi-Fermi approach have been applied by Misner,
Thorne and Wheeler \cite{MTW} for construction the so-called "proper reference
frame" of an accelerating and rotating observer. For an accelerating observer
without rotation the MTW construction corresponds directly to the Synge
approach with its Fermi-Walker transport but for the accelerating and rotating
frame MTW generalized the transport law by adding to evolution equation of the
frame the term proportional to its angular velocity. In any case there are no
Fermi coordinates in these constructions and this was clear for the authors
because they do not claimed any relation of their frame of an accelerating
observer to Fermi coordinates. They refer shortly to Fermi coordinates only at
the end of their analysis in connection to the particular case of the
non-accelerating and non-rotating geodesic observer when Synge prescription
indeed lead to coordinates of Fermi type.
\bibitem {SS2}The Latin indices run through the 4 values 0,1,2,3 and Greek
indices take 3 values 1,2,3. The simple partial derivatives we denote by comma
and covariant derivatives by semicolon. We stress that in general $x^{0}$ has
no obligation to be timelike. One can chose any distribution of signs in
signature of the metric tensor.
\bibitem {SS3}Of course, the particular case for geodesic lines is also
included. If the line (\ref{A1}) is geodesics then functions $f^{\alpha}%
(x^{0})$ cannot be arbitrary but should follow from geodesic equations in
metric $g_{ik}(x)$. Without loss of generality we can consider the metric in
Fermi coordinates (\ref{A36})-(\ref{A38}) having the form $\acute{g}_{ik}%
=\eta_{ik}+O\left(  2\right)  $ where $\eta_{ik}$ is Minkowski tensor. Then
near the line the spacetime is flat and geodesics will be just a straight
line. We can use the Lorentz rotation and shift of the origin of coordinates
(which group do not changes $\eta_{ik}$ and zero value of $\acute{\Gamma}%
$-symbols on line) in order to make this geodesics coincides with coordinate
line of the variable $\acute{x}^{0}$ starting from the origin of the Fermi
coordinate system. This means that equations of line will be $\acute
{x}^{\alpha}=0$ [which means that transformation coefficients $X^{\alpha
}(x^{0})$ in (\ref{A5}) vanish] and all functions $F^{\alpha}\left(  \acute
{x}^{0}\right)  $ in metric (\ref{A36})-(\ref{A38}) disappear. In this case
some part of the equations (\ref{A15}), (\ref{A16}) and (\ref{A18}) will be
equivalent to the geodesic equations for the line $x^{\alpha}=f^{\alpha}%
(x^{0})$ in metric $g_{ik}(x)$ and all other relations will define the rest of
the transformation coefficients in (\ref{A5}). The results will be the same as
in paper \cite{MM} (with the difference that the Riemann tensor in our paper
has defined with opposite sign).
\bibitem {SS4}Expressions (\ref{A84})-(\ref{A87}) follow from the first of the
formulas (14) and from the formulas (74) in the paper \cite{BGJ}. With
approximation up to the second order $O(2)$ the expansion (74) can be easily
inverted, that is Singe's coordinates $X,Y,Z$ can be written in form of
expansion in terms of the Schwarzschild coordinates. This results in expansion
(\ref{A84})-(\ref{A87}) in which we used our notations $m,r_{\ast}%
,\varphi_{\ast}$ for the same constant parameters $M,r_{0},\varphi_{0}$
adopted in paper \cite{BGJ}.
\end{thebibliography}
\end{document}